\documentclass[11pt,twoside]{article}
\usepackage{graphicx}

\begin{document}

\title{Search for Hidden Turbulent Gas through Interstellar  Scintillation}
\author{Moniez M.,$^1$ Ansari R.,$^1$ Habibi F.,$^2$ and Rahvar
  S. $^3$}

\maketitle

$^1$Laboratoire de l'Acc\'{e}l\'{e}rateur Lin\'{e}aire,
{\sc IN2P3-CNRS}, Universit\'e de Paris-Sud, B.P. 34, 91898 Orsay
Cedex, France

$^2$School of Astronomy, Institute for Research in Fundamental Sciences (IPM), PO Box 19395-5531,Tehran, Iran

$^3$Department of Physics, Sharif University of Technology
PO Box 11365-9161, Tehran, Iran
and
Perimeter Institute for Theoretical Physics, 31 Caroline Street North,
Waterloo, Ontario N2L 2y5, Canada \\

\vspace{1.cm}
{\it Talk given in ``Structure and Dynamics of Disk Galaxies'', Petit Jean Mountain, Arkansas
USA, 11-16 august 2013}

\begin{abstract}
Stars twinkle because their light propagates through the atmosphere.
The same phenomenon is expected when the light of remote stars
crosses a Galactic -- disk or halo -- refractive medium
such as a molecular cloud.
We present the promising results of a test performed with the ESO-NTT
and the perspectives of detection.
\end{abstract}
\section{What is interstellar scintillation?}
\begin{figure}[h]
\begin{center}
\includegraphics[width=11.cm]{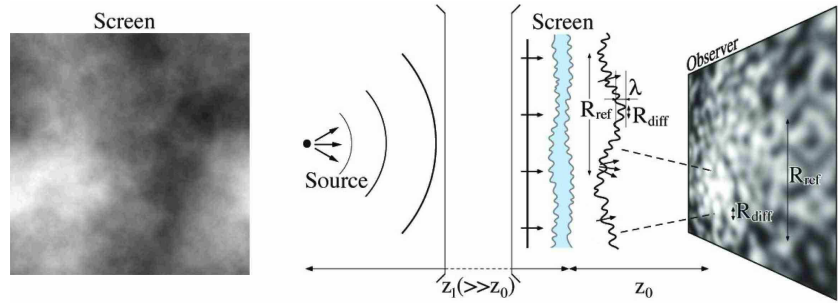}
\end{center}
\vspace{-0.7cm}
\caption[] 
{\it
Left: a 2D stochastic phase screen (grey scale), from
a simulation of gas affected by Kolmogorov-type turbulence.
Right: the illumination pattern from a point source (left) after crossing
the phase screen.
Distorted wavefront produces structures at scales $\sim R_{diff}(\lambda)$
and $R_{ref}(\lambda)$ on the observer's plane.
}
\label{front}
\end{figure}

\begin{figure}[h]
\begin{center}
\includegraphics[width=11cm]{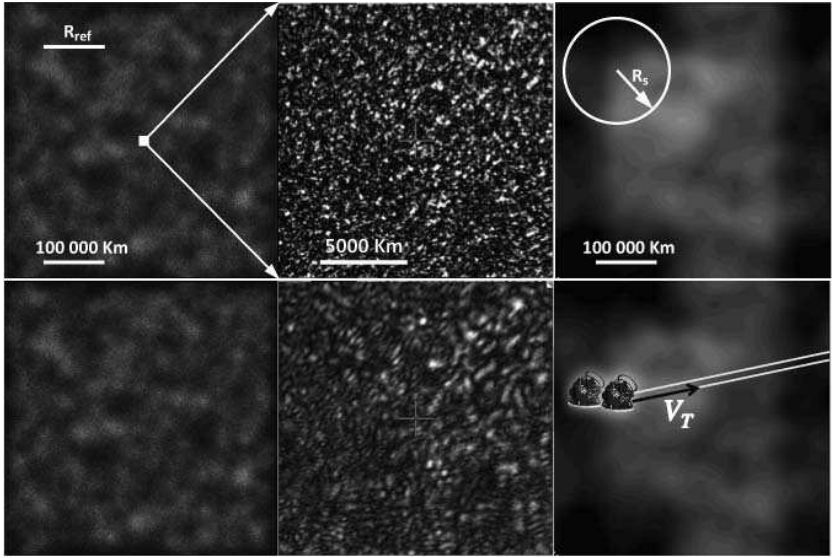}
\end{center}
\caption[] 
{\it
Simulated
illumination maps at $\lambda=2.16\mu m$ on Earth from a point source (up-left)-
and from a K0V star ($r_s=0.85R_{\odot}$, $M_V=5.9$) at $z_0+z_1=1.16\, kpc$ (right).
The refracting cloud is assumed to be at $z_0=160\, pc$
with $R_{diff}(2.16\mu m)=100\, km$.
The circle shows the projection of the stellar disk ($r_s\times z_0/z_1$).
The bottom maps are illuminations in the $K_s$ band
($\lambda_{central}=2.162\mu m$, $\Delta\lambda = 0.275\mu m$).
}
\label{simuscint}
\end{figure}

Refraction through
an inhomogeneous transparent cloud (hereafter called screen)
distorts the wave-front of incident electromagnetic waves
(Fig. \ref{front}) \cite{Moniez};
for a {\it point-like} source,
the intensity in the observer's plane is affected by
interferences which, in the case of stochastic inhomogeneities,
takes on the speckle aspect. Two distance scales characterise this speckle:
\begin{itemize}
\item
The diffraction radius $R_{diff}(\lambda)$ of the screen,
defined as the transverse separation
for which the root mean square of the phase difference at wavelength
$\lambda$ is 1 radian.
\item
The refraction radius $R_{ref}(\lambda)=\lambda z_0/R_{diff}(\lambda)$
where $z_0$ is the distance to the screen.
This is the size, in the observer's plane, of the diffraction spot from a patch
of $R_{diff}(\lambda)$ in the screen's plane.
\end{itemize}
After crossing a fractal cloud described by the Kolmogorov turbulence
law (Fig. \ref{front}, left), the light from a {\it monochromatic
point} source produces an illumination
pattern on Earth made of speckles of size $R_{diff}(\lambda)$ within
larger structures of size $R_{ref}(\lambda)$ (Fig. \ref{front}, right)\cite{simu}.
The illumination pattern from a stellar source of radius $r_s$ is much
less contrasted, since it is
the convolution of the point-source intensity pattern with the projected
intensity profile of the source
(Fig. \ref{simuscint}, right).
%
The cloud, moving with transverse velocity $V_T$
relative to the line of sight, induces stochastic
intensity fluctuations of the illumination at a given point
with the characteristic time scale
\begin{equation}
t_{ref}(\lambda) = \frac{R_{ref}(\lambda)}{V_T} \sim
5.2\, minutes\left[\frac{\lambda}{1\mu m}\right]\left[\frac{z_0}{1\,
    kpc}\right]\left[\frac{R_{diff}(\lambda)}{1000\,
    km}\right]^{-1}\left[\frac{V_T}{100\, km/s}\right]^{-1},
\label{dureescint}
\end{equation}
and modulation index $m_{scint.}=\sigma_I/\bar I$ given by
\begin{equation}
m_{scint.} = 0.035 \, \left[\frac{\lambda}{1 \mu m}\right] \left[\frac{z_0}{1 Kpc}\right]^{-1/6} 
                      \left[\frac{R_{diff}(\lambda)}{1000
                          km}\right]^{-5/6}
                      \left[\frac{r_s/z_1}{R_\odot/10
                          kpc}\right]^{-7/6}.
\label{xparam} 
\end{equation}
%
{\bf Signature of the scintillation signal:}
In addition to the stochasticity, the time scale and
scintillation index, several specificities characterise a scintillation signal.
\begin{itemize}
\item
Chromaticity:
We expect a small variation of the characteristic time scale $t_{ref}(\lambda)$
between the red side of the optical spectrum and the blue side.
\item
Spatial decorrelation:
We expect a decorrelation between the
light-curves observed at different telescope sites, increasing with
their distance.
\item
Correlation between the stellar radius and the modulation index:
Big stars scintillate less
than small stars through the same gaseous structure.
\item
Location:
Extended gas structures should induce (decorrelated)
scintillation of apparently neighboring stars.
\end{itemize}
The first two signatures are probably the strongest ones, since they
point to a propagation effect, which is incompatible with
any type of intrinsic source variability.

{\bf Foreground effects, background to the signal:}
Atmospheric {\it intensity} scintillation is negligible
through a large telescope \cite{dravins}.
Any other atmospheric effect should be easy
to recognize as it is a collective effect.
Asterosismology, granularity of the
stellar surface, spots or eruptions
produce variations of very different amplitudes and time scales.
A rare type of recurrent variable stars exhibit emission
variations at the minute scale, but such objects could be identified
from their spectrum.
\section{Preliminary studies with the NTT}
\begin{figure}[h]
\centering
\parbox{8.3cm}{
\includegraphics[width=6.5cm]{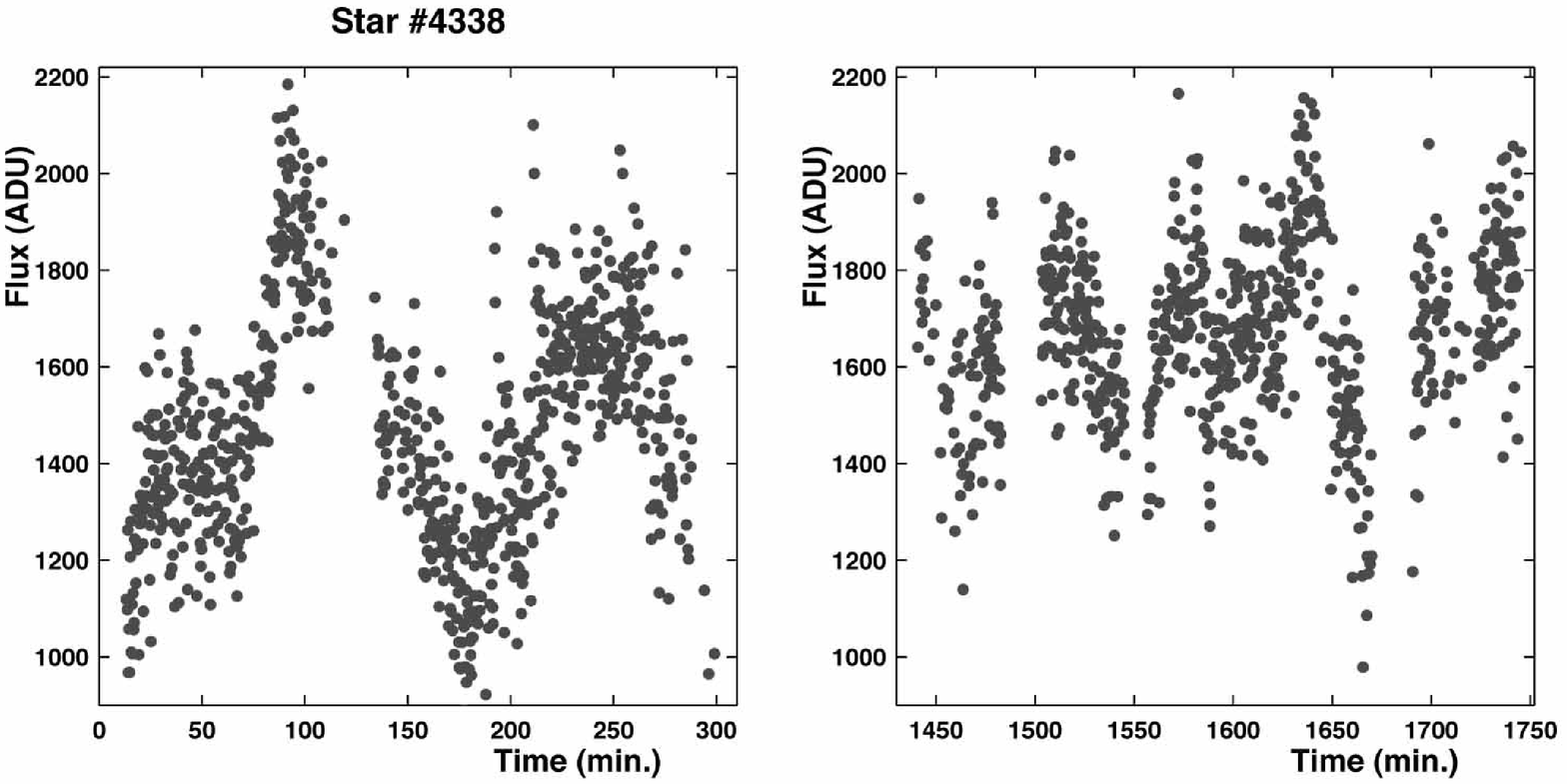}
\caption[]{\it
Light-curves for the two nights of observation (above) and images of the
candidate found toward B68 during low and hight luminosity phases
(right). North is up, East is left.
\label{candidate}}
}
\parbox{4cm}{
\includegraphics[width=3.5cm]{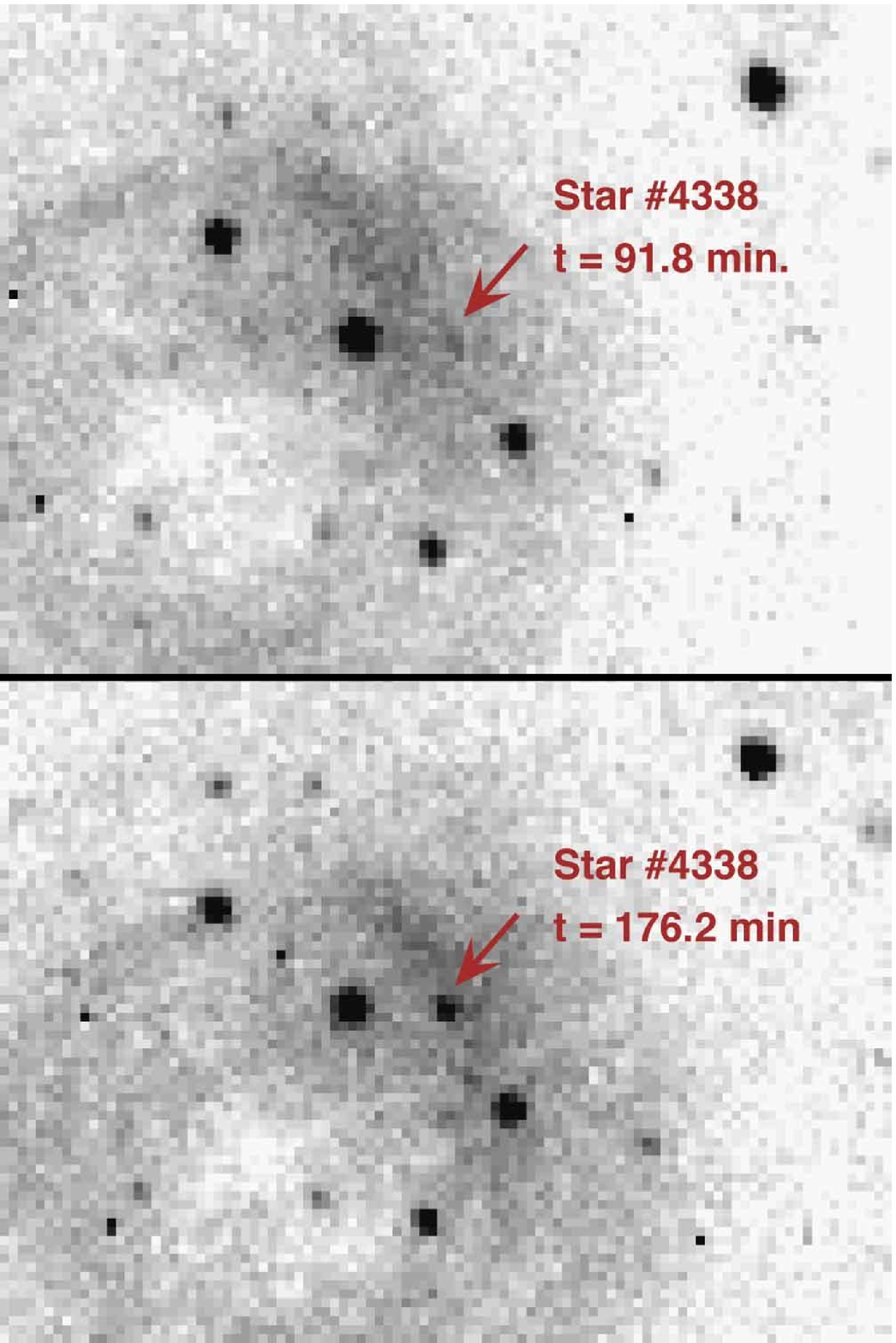}
}
\end{figure}

During two nights of June 2006,
4749 consecutive exposures of ${T_{exp}=10\,s}$
have been taken with the infra-red
SOFI detector in $K_s$ and $J$ through nebulae B68, cb131, Circinus and
towards SMC \cite{resultNTT}.
A candidate has been found towards B68
(Fig. \ref{candidate}), but the poor photometric precision in $K_s$ and
other limitations prevent us from definitive conclusions.
Nevertheless, we can conclude from the rarity of stochastically
fluctuating objects
that there is no significant population of stars that can
mimic scintillation effects, and future searches 
should not be overwhelmed by background of fakes.

\begin{figure}[h]
\centering
\parbox{5.5cm}{
\includegraphics[width=5.5cm,height=5.cm]{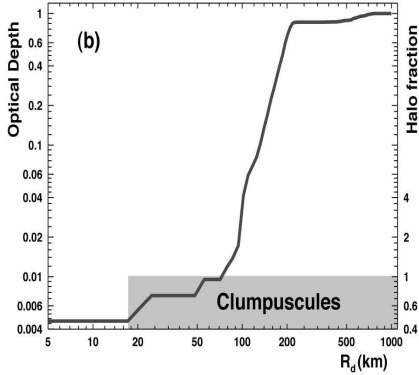}
}
\hspace{0.3cm}
\vspace{-0.3cm}
\parbox{6.0cm}{\caption[]{\it
The $95\%\,CL$ maximum optical depth of structures with $R_{diff}(1.25\mu m)<R_d$
toward the SMC. The right scale gives the maximum contribution of
structures with $R_{diff}(1.25\mu m)<R_d$ to the Galactic halo (in fraction);
the gray zone shows the possible region for the hidden gas
clumpuscules expected from the model of \cite{fractal}.
\label{limits}}}
\end{figure}

From the observed SMC light-curves we also
established upper limits
on invisible gaseous structures as a function of their diffraction radius
(Fig. \ref{limits}). This limit, although not really competitive,
already excludes a major contribution of strongly turbulent gas to the
hidden Galactic matter.
These constraints are currently limited by the statistics
and by the photometric precision.\\
\section{Perspectives}
\begin{figure}[h]
\centering
\includegraphics[width=6.cm]{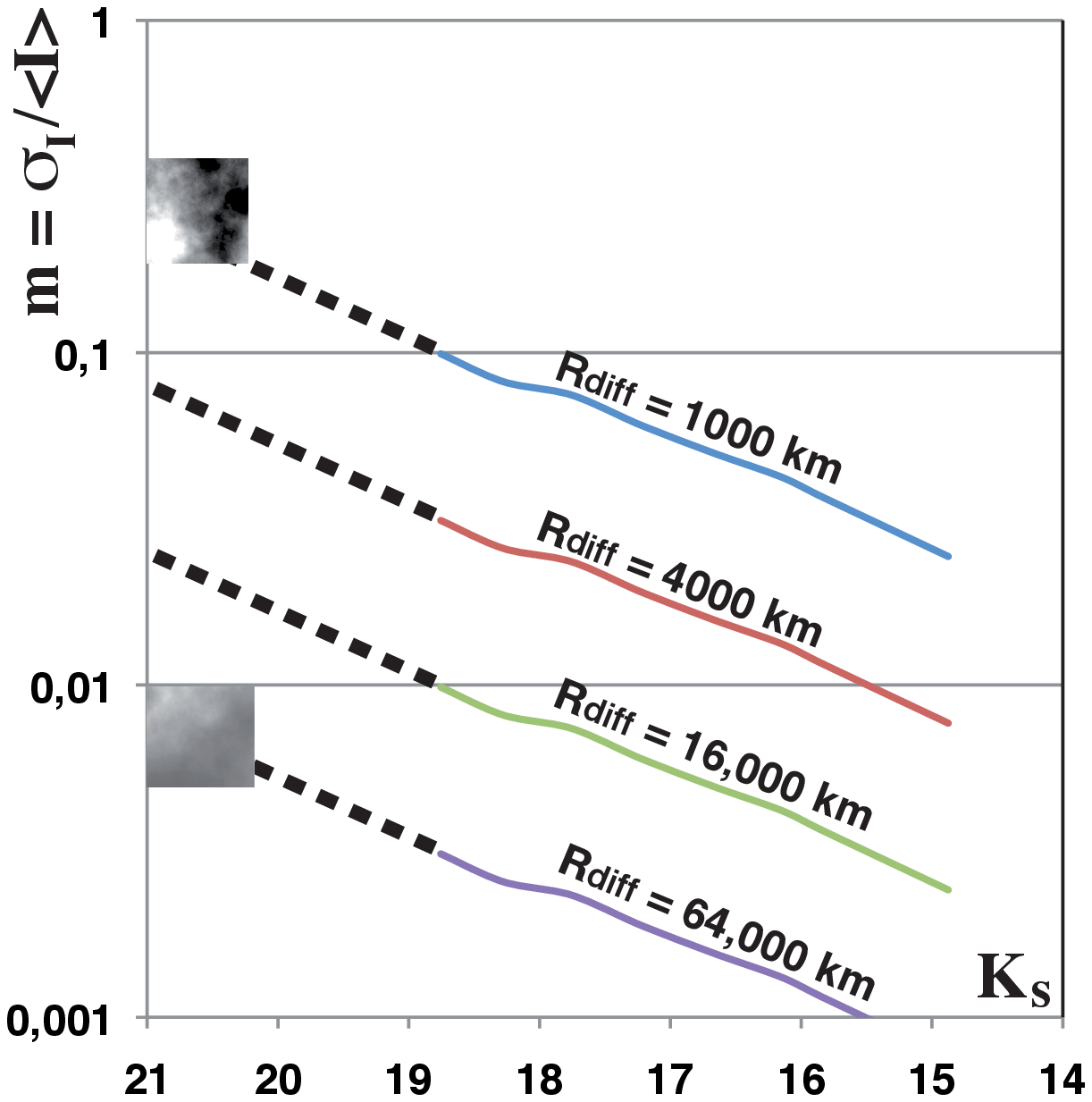}
\includegraphics[width=6.cm]{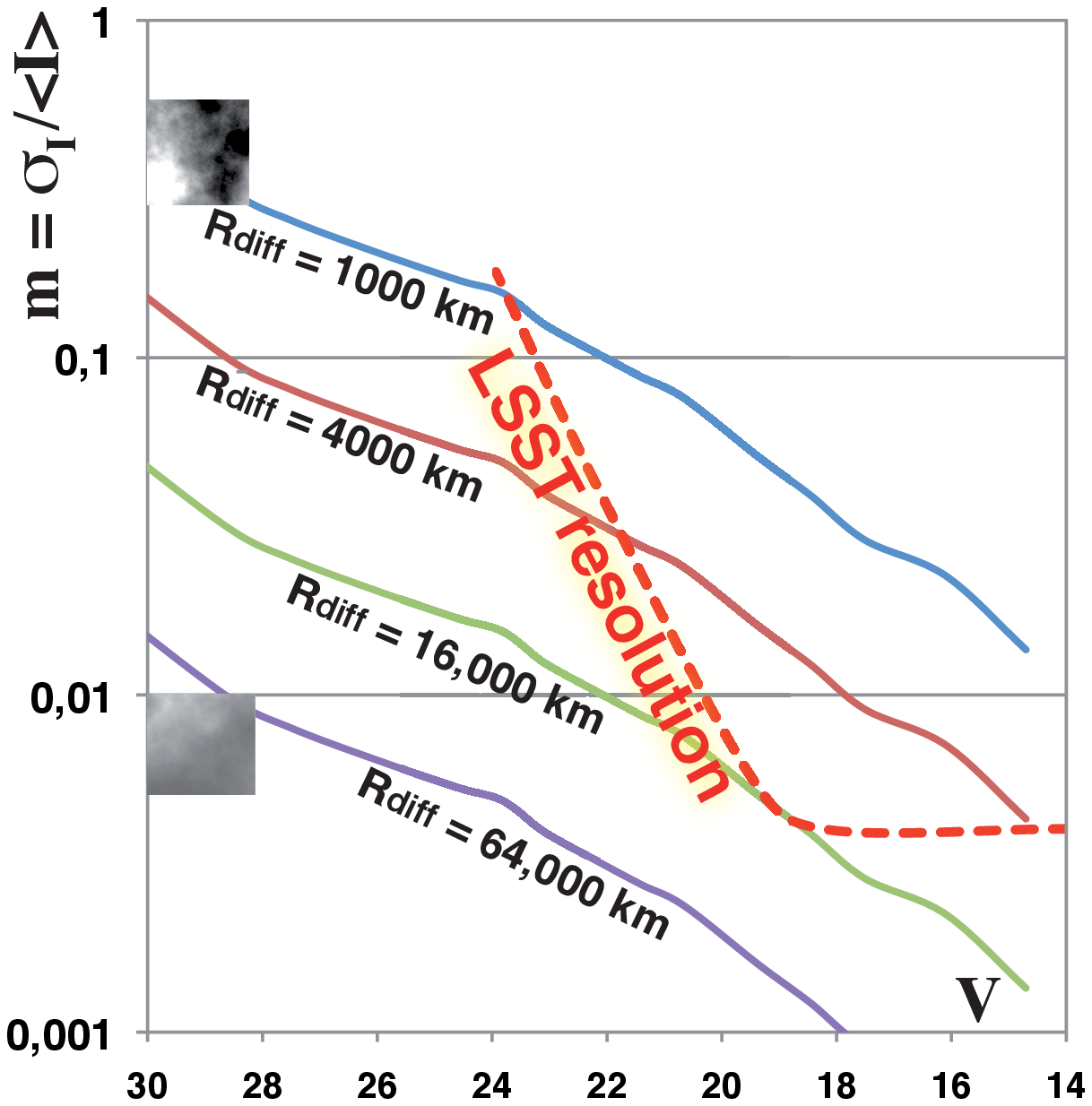}
\caption[]{\it
The expected modulation index {\bf m} as a function of the
source apparent magnitude, for 4 different values of $R_{diff}$.

LEFT: Screen (B68) at $z_0=160pc$, source at $z_1=7Kpc$, observed in
{\bf $K_s$} band.

RIGHT: Screen (invisible halo clumpuscule) at $z_0=1Kpc$, source at
$z_1=55Kpc$ (within LMC), observed in {\bf V}. 
The configurations above the LSST photometric uncertainty curve
(15s exposures) would produce detectable scintillation.
\label{sensitivity}}
\end{figure}

Fig. \ref{sensitivity} allows one to estimate the sensitivity
to turbulent gas when knowing the photometric
performances of an observation system, assuming that the
light-curve sampling is sufficient to observe the time structure
($\Delta t_{sampling}\ll t_{ref}$).
LSST will be an ideal setup for such a search
thanks to the fast readout and to the wide and deep field.

Scintillation signal would provide a new tool to
measure the inhomogeneities and the dynamics of nebulae,
and to probe the molecular hydrogen contribution
to the Milky-Way baryonic hidden matter.

\vspace{-0.5cm}
\begin{thebibliography}{}
\bibitem[Dravins et al. 1998]{dravins}
Dravins, D. {\it et al.}, Pub. of the Ast. Soc. of the Pacific
{\bf 109} (I, II) (1997), {\bf 110} (III) (1998).
\bibitem[Habibi et al. 2011]{resultNTT}
Habibi F., Moniez M., Ansari R., Rahvar S. (2011) A\&A 525, A108.
\bibitem[Habibi et al. 2013]{simu}
Habibi F., Moniez M., Ansari R., Rahvar S. (2013) A\&A 552, A93.
\bibitem[Moniez 2003]{Moniez}
Moniez, M. (2003) A\&A 412, 105.
\bibitem[Pfenniger \& Combes 1994]{fractal}
Pfenniger, D. \& Combes, F. (1994) A\&A 285, 94.
\end{thebibliography}
\end{document}